\documentstyle[amssym,emulateapj]{article} 
\input{psfig} 
\input{epsf}
\newcommand{\minmag}{\raise-2.truept\hbox{\rlap{\hbox{$<$}}\raise 6.truept\hbox
{$>$}}}
\newcommand{\be}{\begin{equation}}
\newcommand{\ee}{\end{equation}}
\newcommand{\ba}{\begin{eqnarray}}
\newcommand{\ea}{\end{eqnarray}}
\newcommand{\brr}{\begin{array}}
\newcommand{\err}{\end{array}}
\newcommand{\bc}{\begin{center}}
\newcommand{\ec}{\end{center}}

\newcommand{\etal}{{et al.}~}

\newcommand{\f}{\frac}

\newcommand{\de}{\delta}

\newcommand{\sigm}{\sigma_{_{\!\!M}}}

\submitted{Draft version \today}
\lefthead{Porciani, Catelan and Lacey}
\righthead{Halo clustering in Lagrangian space}

\begin{document}

\title{How dark matter halos cluster in Lagrangian space}
\author{ 
Cristiano Porciani\altaffilmark{1, 2}, 
Paolo Catelan\altaffilmark{3}
{\tiny AND} 
Cedric Lacey\altaffilmark{3}}
\affil{ \altaffilmark{1} Space Telescope Science Institute, 3700 San 
Martin Drive, Baltimore, MD 21218, USA}
\affil{ \altaffilmark{2} Scuola Internazionale Superiore di Studi 
Avanzati, via Beirut 4, 34014 Trieste, Italy} 
\affil{ \altaffilmark{3} Theoretical Astrophysics Center, Juliane
Maries Vej 30, 2100 Copenhagen \O, Denmark} 

\begin{abstract}
We investigate the clustering of dark matter halos in Lagrangian space
in terms of their two-point correlation function, spanning more than 4
orders of magnitudes of halo masses. Analyzing a set of collisionless
scale-free $128^3$-particle N-body simulations with spectral indices
$n=-2,-1$, we measure the first two Lagrangian bias parameters $b_1$
and $b_2$ relating halo and mass correlations.  We find that the Mo \&
White leading-order formula for $b_1$ describes the clustering of halos
with mass $M \gtrsim M_\star$ quite accurately, where $M_\star$
indicates the characteristic non-linear mass.  Smaller halos turn out
to be less clustered in Lagrangian space than predicted by Mo \& White.
Our findings are consistent with the recent results of Jing for the
clustering of halo populations in Eulerian space, demonstrating that
the discrepancies between the $N$-body and analytical Mo \& White
prediction for the bias exist already in Lagrangian space. This shows
that a more refined theoretical algorithm for selecting halos in the
initial conditions needs to be developed. Finally, we present a very
accurate fitting formula for the linear halo bias factor $b_1$ in
Lagrangian space.
\end{abstract}

% The different journals have different requirements for keywords.  The
% keywords.apj file, found on aas.org in the pubs/aastex-misc directory, 
% contains a list of keywords used with the ApJ and Letters.  These are 
% usually assigned by the editor, but authors may include them in their 
% manuscripts if they wish. 

 
% That's it for the front matter.  On to the main body of the paper.
% We'll only put in tutorial remarks at the beginning of each section
% so you can see entire sections together.

% In the first two sections, you should notice the use of the LaTeX \cite
% command to identify citations.  The citations are tied to the
% reference list via symbolic KEYs.  We have chosen the first three
% characters of the first author's name plus the last two numeral of the
% year of publication.  The corresponding reference has a \bibitem
% command in the reference list below.
%
% Please see the AASTeX manual for a more complete discussion on how to make
% \cite-\bibitem work for you.   

\section{Introduction}

Virialized dark matter halos in the Universe are not distributed in the
same way as the underlying dark matter. This is true whether one takes
the final positions of the halos in Eulerian space or their initial
positions in Lagrangian space (Mo \& White 1996 (MW); Catelan \etal
1998 (CLMP); Catelan, Matarrese \& Porciani 1998; Jing 1998; Sheth \&
Lemson 1998). The numerical tour de force by Jing (1998), who
thoroughly investigated with unmatched accuracy the clustering of dark
matter halos in Eulerian space, demonstrated explicitly that $i)$ the
halo-to-mass bias is independent of the halo separation (at least in
the scale-free case $n=-2$ and in the linear regime); $ii)$ the MW
Eulerian linear bias correctly describes the clustering of halos of
masses $M \!\gtrsim \!M_\star$, but systematically underpredicts it for
any value of the spectral index $n$ if $M \lesssim M_\star$, where
$M_\star$ is the typical non-linear mass. However, it is impossible to
understand solely on the basis of the Eulerian investigation whether
the discrepancies between the numerical results and the analytical MW
predictions are due to a failure of the algorithm for identifying the
halo positions in Lagrangian space, or to the effects of non-linear
shear dynamics (not accounted for in the original MW approach) on the
mapping of halo positions from Lagrangian to Eulerian space, or to a
combination of the two.

In this Letter, we employ scale-free collisionless $N$-body simulations
(i.e. with density parameter $\Omega=1$ and initial power spectra
$P(k)\propto k^n$) to investigate the clustering of dark matter halos
in Lagrangian space in terms of their two-point correlation function,
spanning more than 4 orders of magnitudes in halo masses. We compare
the halo correlation function to the correlation of the underlying dark
matter for $n = -2, -1$.

We find that the theoretical `underclustering' reported by Jing for
masses $M \lesssim M_\star$ is already present in the initial
conditions as well (but as `overclustering', since the first-order bias
is negative for small masses), and cannot be due exclusively to the
subsequent non-linear effects of the shear dynamics acting on small
scales and unaccounted for in the original spherical collapse model of
MW. We then argue that the standard Press-Schechter approach
(`extended' or not), on which Mo \& White (1996) and Catelan \etal
(1998) based their speculations, is inadequate for identifying the
locations of halos in the initial conditions, and a major effort should
be devoted to fin- ding an improved algorithm. \S$\,2$ reports the
details of the present investigation.  Finally, an accurate fitting
formula for the bias $b_1$ in Lagrangian space is given, which should
be considered as the Lagrangian version of Jing's fitting formula for
the Eulerian case. \S$\,3$ contains our conclusions.

\section{Halo clustering from $N$-body simulations}

\subsection{Simulations and Halos}
The simulations used here are similar to those of Lacey \& Cole (1994),
who used them to test halo merging histories.  They were performed
using the ${\rm AP}^3{\rm M}$ code of Couchman (1991) with $128^3$
particles.  The force-softening used was $L/1280$, with $L$ the size of
the periodic box.  Initial positions and velocities were generated by
displacing particles from a uniform $128^3$ grid according to the
Zel'dovich approximation, assuming an initial scale-free power spectrum
and Gaussian statistics.  We considered 4 realizations for 2 different
spectral indices $n=-2, -1$.  For each simulation we recorded, for many
epochs, positions and velocities of all particles.  The output times
were chosen so that $M_{\star}$ increased by a factor $\sqrt{2}$
between subsequent output times.

For each output time we selected dark matter halos in the simulations
employing the `friends-of-friends' group finder with a linking length
equal to 20\% of the mean interparticle distance (e.g. Davis \etal
1985). We checked that results obtained using the spherical overdensity
group finder (Lacey \& Cole 1994) are essentially identical.  We
excluded halos containing less than 20 particles or more than 20000
particles.  We moved all the particles belonging to a given halo back
to their initial (Lagrangian) positions, then computed the position of
their centre of mass.  We used the latter as the `halo position' in
Lagrangian space.  In such a way, for each output time, we constructed
a catalogue of halos indicating their mass and position in Lagrangian
space.

\subsection{Halo correlation function in Lagrangian space}
We computed the mean correlation function $\bar \xi_h$ between halos in
a given mass interval, where a bar denotes mass averaged quantities.
Self-similar scaling allowed us to combine data from different output
times in order to reduce the Poisson fluctuations due to the finite
number of halos within the box.  We considered every output time
contai- ning more than 100 halos in the same mass interval, and we
binned the distributions of halo separations in units of $r/R_\star$,
where $R_\star^3 \propto M_\star$. Finally, we computed the halo
correlation function using the estimator
\be \langle \bar \xi_h \rangle = \Big\{\!\sum _i
N_i[r/R_\star(z_i)]\!\Big\}/\Big\{\!\sum _i N_i^{\rm Poi}
[r/R_\star(z_i)]\!\Big\} - 1\;, 
\ee 
where the index $i$ runs over different output times $z_i$ of the same
simulation, $N_i(r/R_\star)$ is the number of halo pairs in the $i$-th
output, and $N_i^{\rm Poi}(r/R_\star)$ is the corresponding quantity
for a Poisson process with the same number density. The average symbol
$\langle\,\rangle$ is introduced since we considered information coming
from the different temporal outputs.  In this way, for each mass
interval, we collected four realizations of the Lagrangian correlation
function of dark matter halos.

This procedure allows us to achieve two goals: {\it i)} to reduce
statistical fluctuations by increasing the number of halo pairs; {\it
ii)} to extend the mass interval and the range of halo separations that
may be sampled.  In fact, the box size $L$ and the minimum halo mass in
the simulation are fixed, while $R_\star$ and $M_\star$ increase with
time.  This means that the correlation function for halos with $M\gg
M_\star$ is measured mainly from the early output times, while that for
halos less massive than $M_\star$ comes mostly from later output times.
In order to simulate bootstrap resampling, we assigned as the standard
error the Poisson errorbar multiplied by a factor $\sqrt{3}$ (Mo, Jing
\& B\"orner 1992).
  
\subsection{Lagrangian bias parameters}
Now we want to test whether the Lagrangian halo correlation $\xi_h$ is
related that of the mass ($\xi_m$) through a relation (see Catelan et
al. 1998; Porciani et al. 1998 and references therein)
\be
\xi_{h}\Big(\f{r}{R_\star}\Big)=b_1^2\,\xi_m\Big(\f{r}{R_\star}\Big)
+\f{b_2^2}{2}\,\xi_m^2\Big(\f{r}{R_\star}\Big)+\dots \;,
\label{xih}
\ee
where the symbol $b_i$ denotes the $i$-th Lagrangian bias factor, and
$\xi_m$ is calculated according to linear theory.  For instance, the
`extended' Press-Schechter approach 
%%%%%%%%%%%%%%%%%%%%%%%%%%%%%%%%%%%%%%%%%%%%%%%%%%%%%%%%%%%%%%%%%%%%
\centerline{ 
\vbox{ 
\hbox{
\psfig{figure=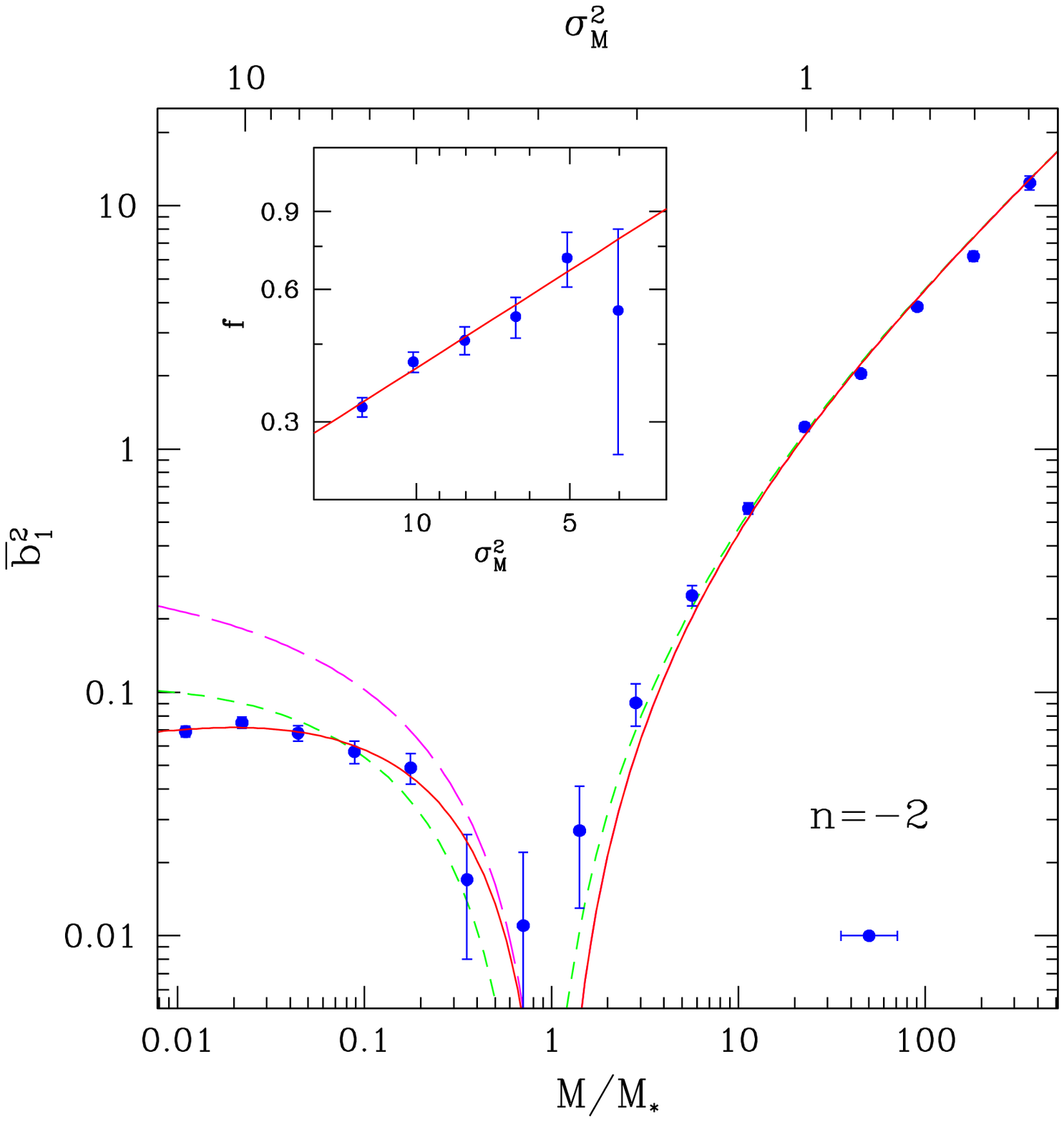,bbllx=32pt,bblly=225pt,bburx=513pt,bbury=684pt,clip=t,height=7.18cm,width=8.0cm}
} 
\hbox{
\psfig{figure=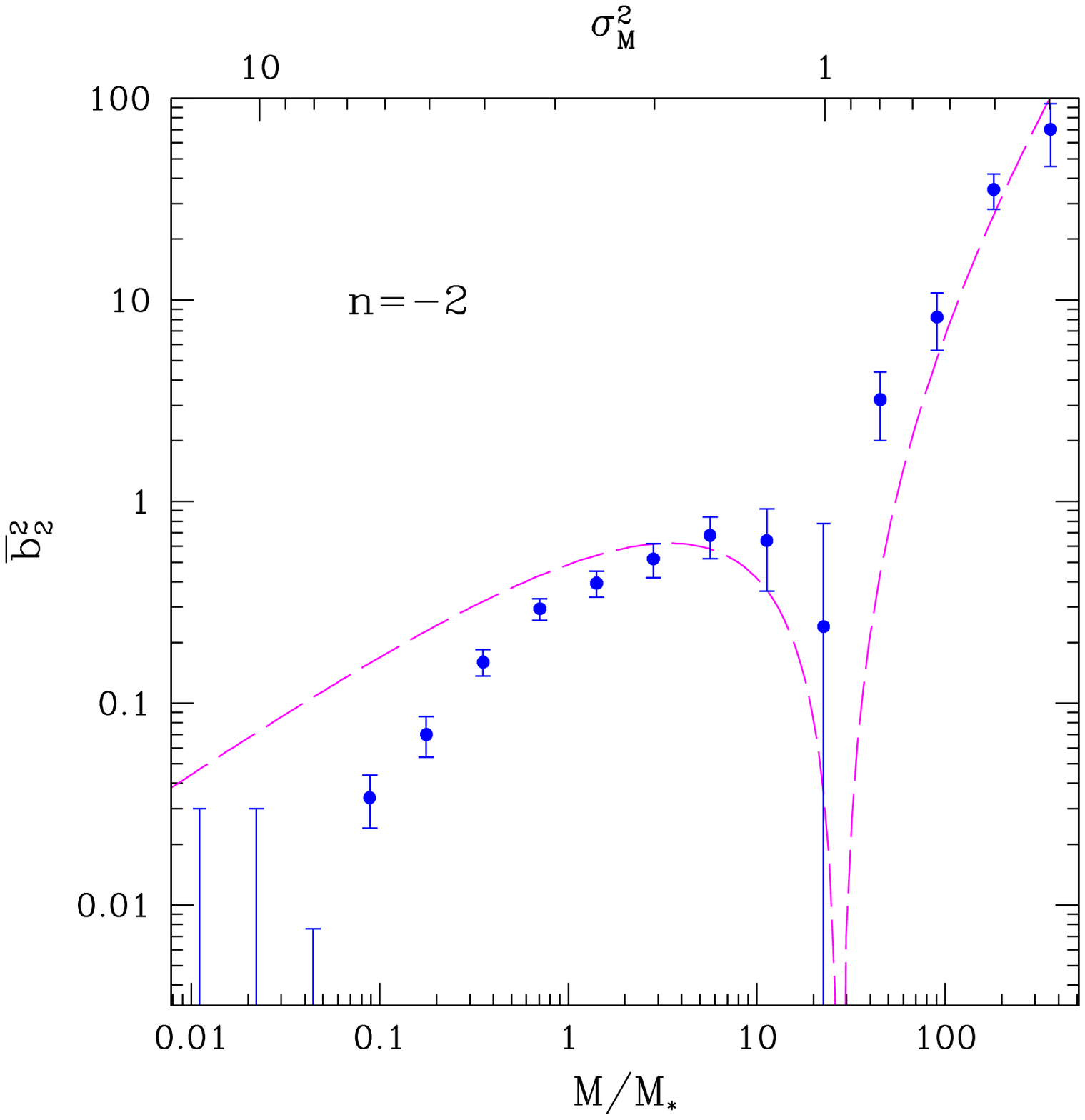,bbllx=32pt,bblly=180pt,bburx=513pt,bbury=642pt,clip=t,height=8.0cm,width=8.0cm}
} } } 
{\footnotesize{Fig. 1.} --- {\it Dots:} Best fit values of $\bar
b_1^2$ (top panel) and $\bar b_2^2$ (bottom panel) as a function of
variance $\sigm^2$ and halo mass on the same scale for $n=-2$.
Errorbars denote one-sigma uncertainties. {\it Long-dashed line:} MW and CLMP
Lagrangian bias with $\de_c=1.686$. {\it Short-dashed line:} Jing's
fit to the Eulerian bias minus 1. {\it Continuous line:} our fitting
formula for $b_1$ given in the main text.  Both the theoretical
predictions and the fitting formula are averaged over the
Press-Schechter mass function. Note how the halo correlation remains
non-zero at $M=M_\star$ through the effect of $b_2$, even though $b_1$
vanishes there.  The horizontal errorbar in the bottom-right corner
of top panel indicates the mass intervals considered. The box inside
the top panel shows the best fit correction $f(\sigm)$ to the MW
formula for $b_1$, when $M<M_\star$: $\log f=(0.34 \pm 0.11) - (0.73
\pm 0.12) \log \sigm^2$.}
\vspace{0.4cm}
%%%%%%%%%%%%%%%%%%%%%%%%%%%%%%%%%%%%%%%%%%%%%%%%%%%%%%%%%%%%%%%%%%%%%%

\noindent (Bond et al. 1991) leads to the MW and CLMP expressions for
$b_1$ and $b_2$, namely $b_1=\de_c/\sigm^2-1/\de_c$ and
$b_2=(\de_c^2/\sigm^2-3)/\sigm^2$, where $\sigm^2$ indicates the mass
variance on scale $M$ (Cole \& Kaiser 1989; Mo \& White 1996; Mo, Jing
\& White 1997).  Note that, by definition, $\sigm=\de_c$ for
$M=M_\star$.  To test this biasing model against simulations, we have
to consider a finite range of halo masses.  Eq.(\ref{xih}) then implies
$\bar \xi_h=\bar b_1^2 \xi_m+ \f{1}{2}\bar b_2^2 \xi_m^2+\dots$, where
$\bar b_k$ is the mean of $b_k$ in the mass interval, weighted by the
mass
%%%%%%%%%%%%%%%%%%%%%%%%%%%%%%%%%%%%%%%%%%%%%%%%%%%%%%%%%%%%%%%%%%%%%%%%%%%
\centerline{ \vbox{ \hbox{
\psfig{figure=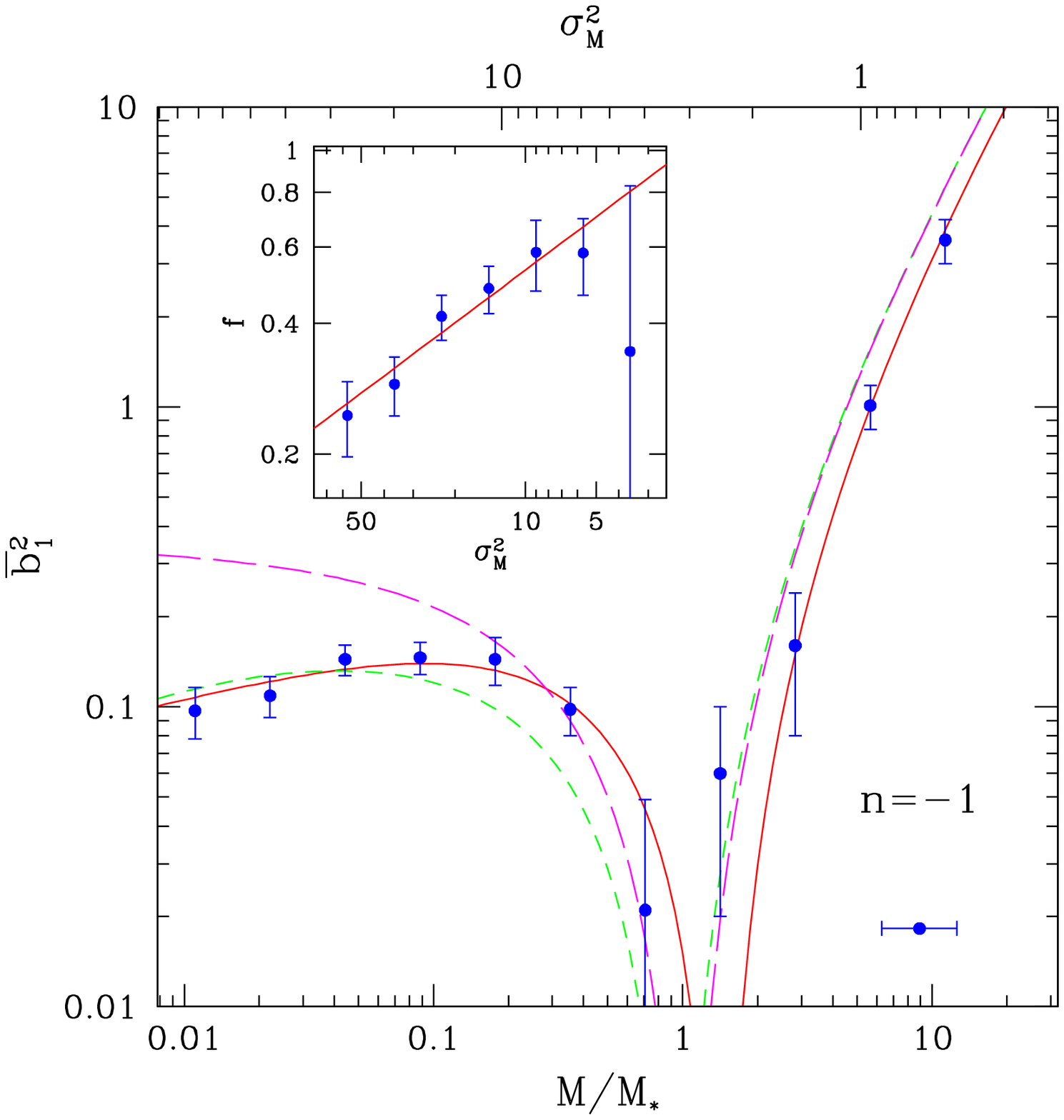,bbllx=32pt,bblly=225pt,bburx=513pt,bbury=684pt,clip=t,height=7.18cm,width=8.0cm}
} \hbox{
\psfig{figure=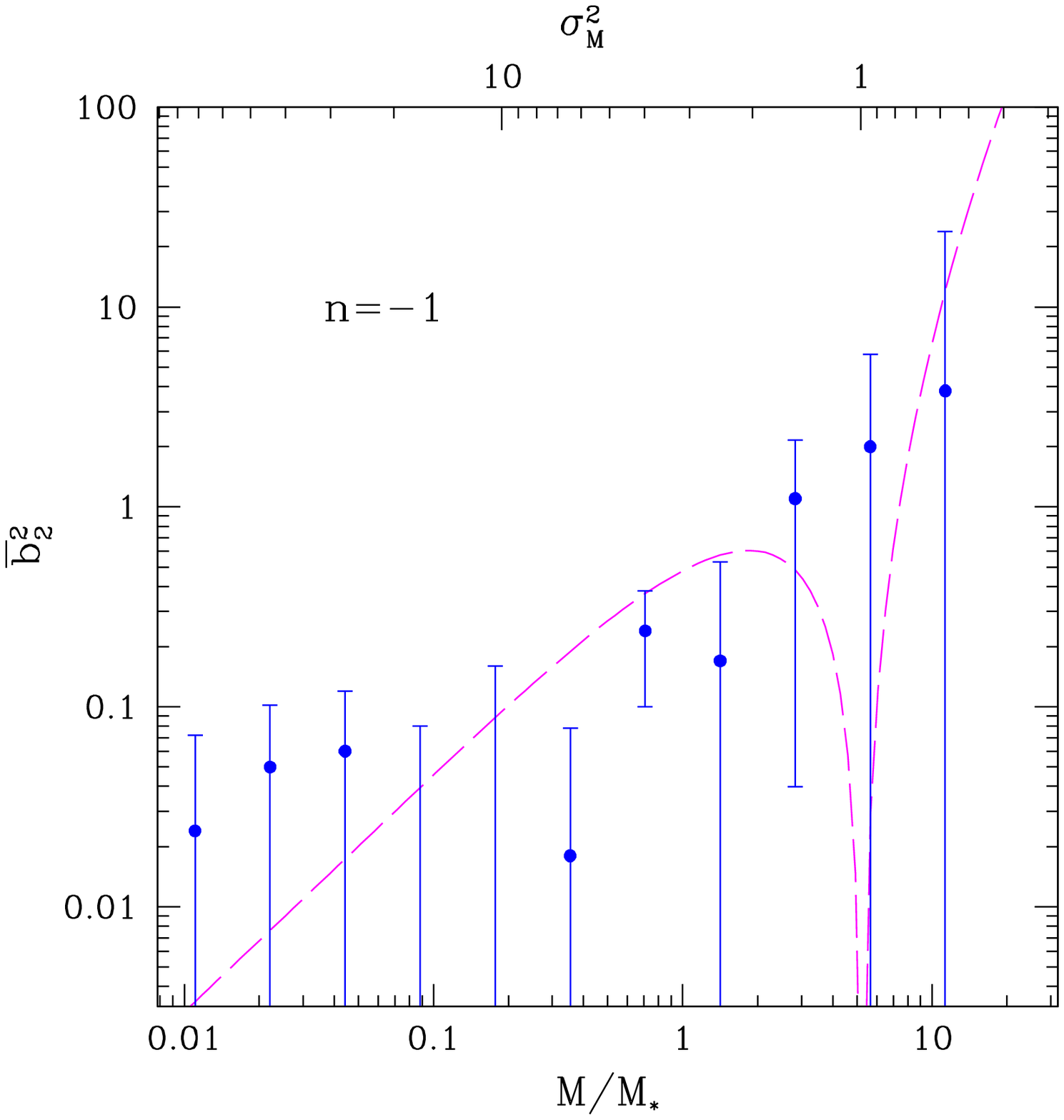,bbllx=32pt,bblly=180pt,bburx=513pt,bbury=642pt,clip=t,height=8.0cm,width=8.0cm}
} } } {\footnotesize{Fig. 2.} --- As in Fig. 1 but for $n=-1$.
In this case, the best fit for $f$ is: $\log f=
(0.13 \pm 0.13) - (0.41 \pm 0.10) \log \sigm^2$}
\vspace{0.4cm}
%%%%%%%%%%%%%%%%%%%%%%%%%%%%%%%%%%%%%%%%%%%%%%%%%%%%%%%%%%%%%%%%%%%%%%%%

\noindent function.  In order to obtain statistically reliable values
for $\bar b_1$ and $\bar b_2$, we compute the averages $\langle \xi_m
\rangle$ and $\langle \xi_m^2 \rangle$ {\it directly} from the initial
conditions of the simulation, following the same averaging procedure we
used for the halo correlations (in this case it corresponds to
averaging over output times using the volume of the $r/R_\star$ bin as
a weight).  Note that, after averaging, $\langle \xi_m^2 \rangle$
generally differs from $\langle \xi_m \rangle^2$, so that
eq.(\ref{xih}) implies $\langle \bar \xi_h\rangle =\bar b_1^2 \langle
\xi_m \rangle + \f{1}{2}\bar b_2^2 \langle \xi_m^2 \rangle +\dots
$. This avera- ging procedure is important to account for the
finiteness of the box: in this way, the lack of any Fourier components
of the density would be equally experienced by both mass and halo
distributions.  Finally, to check the reliability of eq.(\ref{xih}) we
apply the $\chi^2$ test, assuming that every $\langle \bar
\xi_h\rangle$ is normally distributed around a mean value.

Halos are objects of finite size and we expect exclusion effects to
dominate $\xi_h$ at separations of the order of the halo Lagrangian
size $R_L$.  In a sample composed of identical spherical halos, spatial
exclusiveness implies $\xi_h=-1$ for $r<2 R_L$ and a strong
compensating positive correlation at $r \gtrsim 2 R_L$. The magnitude
of this positive correlation is related to the volume fraction occupied
by the spheres, since $\int_0^\infty dr\, r^2\, \xi_h(r)=0$.  Inspired
by this, we decided to exclude from the $\chi^2$ test all data
corresponding to separations $r \lesssim 2R_{\rm max}$, where $R_{\rm
max}=R_\star (M_{\rm max}/M_\star)^{1/3}$ is the characteristic
Lagrangian radius of the largest halo in the mass interval considered;
the exact value of the mini- mum separation considered has been
determined on a case by case basis, checking for the stability of our
results with increasing the number of data points at small separations.
We avoided averaging between different realizations of the same power
spectrum: the entire data set of up to 4 values of $\langle \bar\xi_h
\rangle $ for each $r/R_\star$ interval is considered in the $\chi^2$
test.

Minimization of $\chi^2$ with non-holonomic constraints $\bar b_1^2
\geq 0$ and $\bar b_2^2 \geq 0$ is used to obtain estimates of the bias
parameters.  The resulting $\chi^2_{\rm min}$ shows that the {\it a
priori} bias hypothesis--the bias relation in eq.(\ref{xih})--is
acceptable to a 95\% confidence level over a wide interval of halo
masses.  However, for $n=-2$, samples with $M>16 M_\ast$ are
inconsistent with eq.(\ref{xih}) at the 99\% confidence level.  This is
caused by the presence of noticeable differences (especially at very
large separations) between the halo correlation functions extracted
from different realizations.  Indeed, very massive halos are
exponentially rare and more affected by statistical fluctuations.  On
the other hand, we could also have underestimated the uncertainty in
computing $\langle\xi_h \rangle$.

In our approach a small $\chi^2$ is found only if eq.(\ref{xih}) with
the {\it same} values for $\bar b_1$ and $\bar b_2$ accurately
describes {\it every} clustering realization, for a given $n$. We
accounted for both the dispersion between realizations (since data from
all the realizations enter the $\chi^2$ test ) and the uncertainty of
the single data points (since the residuals between fit and data are
divided by the simulated bootstrap errors).  This approach is different
and much more severe than first averaging $\xi_h$ and $\xi_m$ over the
realizations and finally fitting the data using a least-squares method.
In this case, a larger dispersion between realizations creates larger
errorbars for the averaged quantities, thus making $\chi^2$ smaller.
In Jing (1998) no statistical test is performed: the bias relation at
large separations, where $\xi_m \ll 1$, is assumed to be linear.  For
each separation and realization, the squared Eulerian bias is estimated
by computing $\xi_h/\xi_m$.  These va- lues are then averaged over
different realizations and the estimated dispersion of the mean value
is taken as the errorbar.  Finally, by considering different
separations, it is checked whether the averaged biases have consistent
va- lues, within their statistical uncertainties.

Figures 1 and 2 show the best-fit results for $\bar b_1^2$ and $\bar
b_2^2$.  The errorbars are obtained by projecting along the $b$-axes
the contours of constant $\chi^2$ at $\chi^2_{\rm min}+1$.  Concerning
$\bar b_1^2$, the comparison against theoretical predictions shows a
quantitatively good agreement for $M>M_\star$.  However, the square
bias parameter of lower mass halos is significantly smaller than
predicted by Mo \& White (meaning that the bias $\bar b_1$ is {\it less
negative} than MW value).  Findings of this kind have been published by
Jing (1998), who investigated the Eulerian halo clustering in the
quasi-linear regime, giving an accurate fitting formula for the
Eulerian analogue of $b_1$.  Our result shows incontrovertably that the
effects discove- red by Jing are already present in Lagrangian space.

There is no rigorous way to apply Jing's bias fitting formula in
Lagrangian space, since the non-local dyna- mics of the mass density
field enters the transformation.  However, as a first approach, and
only to check the order of magnitude of the effect, we can follow Mo \&
White (1996) in assuming a spherical evolution of the coarse-grained
mass density field, to map from Lagrangian to Eulerian space.  In this
case, the first Eulerian bias term is given by $1+b_1$.  Jing's
formula, $b_1 \equiv b_{\rm Jing}-1$, turns out to be a reasonably good
description of our data, sho- wing that the `underclustering' of small
halos in Eulerian space detected by Jing is correspondingly already
present in their Lagrangian clustering; the intervening dynamics is
approximately well accounted for by the standard mapping from
Lagrangian to Eulerian space in laminar regime. We elaborate below our
own fitting formula for the linear Lagrangian bias, which turns out to
be accurate to 10\% over the entire mass range investigated (with the
exception of the zero crossing region for $b_1$).

Our results for $b_2$ are of course less conclusive than for $b_1$,
mainly because of the larger uncertainties.  However, they confirm the
information extracted considering $b_1$.  Quantitatively, the CLMP
expression for $b_2$ gives good estimate for halos with $M > M_\star$,
while it overpredicts the numerical outcome for $b_2^2$ when
$M<M_\star$.  For $M \sim M_\star$, our results are a factor of $\sim
1.4$ smaller than predicted by the CLMP formula.  This is the range in
which our method is best suited to compute $b_2$, since $b_1$ vanishes
and the halo correlation function is proportional to $\xi_m^2$.  No
firm conclusion can be drawn for $n=-1$, since practically the entire
data set only allows one to set an upper limit on $\bar b_2^2$. It is
not unlikely that, for the less negative spectral indices, bigger
simulations are better suited to quantify the halo clustering up to
second-order biasing, above all in Lagrangian space.  It will be
worthwhile to address this point in a future work.

\subsection {Accurate fitting formula for $b_1$}
Our results for $b_1$ can be accurately parametrized by introducing a
mass-dependent, multiplicative correction to the MW formula, namely
\be
b_1^2(\sigm)=f(\sigm)\,(\delta_c/\sigm^2-1/\delta_c)^2\;.
\ee
For $n=-2$, the original Mo \& White formula (i.e. $f\equiv 1$) with
$\de_c=1.686$, as suggested by the spherical collapse model, describes
the data for $M>M_\star$ to within 10\% accuracy.  Smaller masses
instead require $f=2.19/\sigm^{1.46}$.  For $n=-1$ and $M>M_\ast$ our
numerical data are very well described by the MW formula for $b_1$ with
a lower collapse threshold $\de_c\simeq 1.52$. In this case, for
$M<M_\ast$ we obtain $f=1.35/\sigm^{0.82}$.  These fitting formulae are
extremely accurate in describing our data set.  Their simple power-law
behavior encourages further theoretical investigation. For instance,
with data for more than two spectral indices $n$, one may attempt to
fit the dependence of $b_1$ on $n$.

\section{Discussion and conclusions}

Employing $128^3$-body scale-free simulations, we analyzed the
clustering of dark matter halos in Lagrangian space. The main results
of this investigation can be summarized as follows: $i)$ assuming a
correlation model as in eq.(2), the first two Lagrangian bias factors
$b_1$ and $b_2$ are strongly mass-dependent over the 4 orders of
magnitude in mass investigated; $ii)$ the clustering of halos with mass
above the non-linear mass, $M \gtrsim M_\star$, is fairly well
described by the MW formula for the linear Lagrangian bias, both for
$n=-1$ and $n=-2$; $iii)$ halos with non-linear masses $M \lesssim
M_\star $ are less clustered (have a smaller correlation amplitude)
than what the leading order Lagrangian bias of Mo \& White would
predict.

When these results are combined with the ones recently obtained by Jing
(1998) about the clustering of halos in Eulerian space, we can
disentangle the question of whether the discrepancies between Jing's
numerical results and the Mo \& White theoretical predictions are
mainly due to the effects of the non-linear shear dynamics, effective
on smaller scales, or to a possible failure of the halo selection
algorithm in the initial conditions--a question actually left unsolved
by Jing. Clearly, since, as we showed, the {\it same} effects
discovered by Jing are essentially already present in Lagrangian space,
our investigation suggests that it is time to improve on the classical
Press-Schechter algorithm for identifying halos in Lagrangian space:
since it assumes spherically symmetric collapse, it is not surprising
that it fails in correctly counting the small halo masses, where
departures from spherical collapse can be cosmologically relevant.  The
failure of the Mo \& White formula for the Lagrangian and Eulerian bias
is presumably related to the departure of the halo mass function from
the Press-Schechter form at low masses, e.g. Lacey \& Cole (1994).

Finally, we derived a fitting formula for the linear Lagrangian bias
$b_1$ that is relevant for accurately predicting the clustering of dark
matter halos, above all in the low-mass tail. Our fitting formula can
be considered as the Lagrangian equivalent of eq.(3) in Jing
(1998). These results are highly relevant for predicting the clustering
of low luminosity galaxies, most of which lie in lower-mass halos
(e.g. Baugh \etal 1998). Modelling of galaxy clustering found in
present and future galaxy redshift surveys will provide an important
application of the present results.

\vspace{0.2cm} CP acknowledges the hospitality of TAC during the spring
of 1997, when this work started. PC warmly thanks Sergio Gelato for
insightful discussions. We all thank Sabino Matarrese for sharing his
ideas on the matter, dark and not. CP thanks support from NASA
ATP-NAG5-4236 grant and from Italian MURST. PC and CL have been
supported by the Danish NRF at TAC.

\end{document}